\def \jobname {proc}
\documentclass[
    ,final            % use final for the camera ready runs
  ,numberedheadings % uncomment this option for numbered sections
  ]
  {aipproc}

\layoutstyle{8x11single}

\begin{document}

\title{Gauge field topology and the hadron spectrum}

%<Replace this text with PACS numbers; choose from this list:
%                \texttt{http://www.aip..org/pacs/index.html}>}

\author{Michael Creutz}{
  address={Brookhaven National Laboratory}
}

\begin{abstract}
Topologically non-trivial gauge field configurations are an
interesting aspect of non-abelian gauge theories. These become
particularly important upon quantizing the theory, especially through
their effect on the pseudo-scalar spectrum. These effects are closely
tied to chiral anomalies and the possibility of CP violation in the
strong interactions. 
\end{abstract}

\maketitle

%%%%%%%%%%%%%%%%%%%%%%%%%%%%%%%%%%%%%%%%%%%%
%% MAINMATTER
%%%%%%%%%%%%%%%%%%%%%%%%%%%%%%%%%%%%%%%%%%%%

%%%%%%%%%%%%%%%%%%%%%
% Mike's header for slides in TeX
\input epsf
% \input colordvi  

% for a bulleted item
\def \li {\par\hskip .2in {$\bullet$}\hskip .1 in }
\def \lli {\par\hskip .4in {$\bullet$}\hskip .1 in }

\def \hi {\medskip}
\def \topic #1{\par\centerline {#1}}

% \fitframe draws a box around its contents (from Arvind)
% #1=line thickness, #2=inner space, #3=contents
\def\fitframe #1#2#3{\vbox{\hrule height#1pt
 \hbox{\vrule width#1pt\kern #2pt
 \vbox{\kern #2pt\hbox{#3}\kern #2pt}
 \kern #2pt\vrule width#1pt}
 \hrule height0pt depth#1pt}}

% to comment out parts of a TeX file
\long \def \blockcomment #1\endcomment{}

\def \nextslide{\par\hrule\par}

\section{introduction}

It has long been recoginized that topology plays a fascinating role in
the theory of gauge fields.  This appears already at the level the
classical theory, with topology at the heart of a multitude of
nontrivial exact solutions to the Yang-Mills field equations.  The
quantum theory inherits many interesting features through topological
excitations appearing in the path integral.  Topology is also
intimately connected with the behavior of fermion fields through the
index theorem.  This directly leads to the the well known anomalies in
the chiral symmetries of the theory, which are crucial to understnding
certain aspects of the spectrum of the pseudoscalar mesons.

This presentation is meant to be an elementary overview of how this
physics fits in with our understanding of QCD.  Much of this material
is adapted from the more extensive review in
Ref.~\cite{Creutz:2011hy}.  I begin in section \ref{classicalsec} with
a brief discussion of how topology becomes relevant to classical gauge
theories through boundary conditions.  Section \ref{quantum} goes on
to discuss some issues that arise in the quantum theory and their
implications for the continuum limit of the lattice theory.  I then
turn in section \ref{index} to how topology becomes particularly
crucial through its effect on quark fields via the index theorem.
This is important to understanding chiral symmetry and the
pseudo-scalar spectrum, as will be discussed in
sections \ref{pseudoscalars} and \ref{isosec}.
Section \ref{latticesec} explores some unresolved issues that arise on
implementing these ideas with a lattice cutoff.  Section \ref{summary}
provides a brief recap of the earlier sections.

\section{Classical gauge fields}
\label{classicalsec}

One path towards understanding how topology enters a gauge theory is through
boundary conditions.  It is natural to impose that the gauge field
tensor $F_{\mu\nu}$ vanishes at spatial infinity.  This alone,
however, is not sufficient to say that the vector potential $A_\mu$
also vanishes there.  Indeed the only implication is that the
potential goes to the form of a pure gauge; i.e. as
$|x_\mu|\rightarrow\infty$ this requires
\begin{equation}
A_\mu \rightarrow -ih^\dagger \partial_\mu h
\end{equation}
where $h(x)$ is an element of the gauge group.

Now formally spatial infinity is a sphere $S_3$ and the gauge group
can also contain spheres.  For example, a general $SU(2)$ element can
be written in the form $h=a_0+\vec a\cdot\vec\sigma$ where
$a_\mu^2=1.$ That is, the group itself is an $S_3$.  Higher groups
generally contain multiple $SU(2)$ subgroups, to which the following
discussion can be applied.  The basic issue is that the gauge function
$h(x)$ can wrap non-trivially around the group as $x$ surrounds
spatial infinity.  Such mappings cannot be smoothly deformed onto each
other.  The space of all gauge fields where $F_{\mu\nu}$ vanishes at
infinity then divides into topological ``sectors'' depending on how
many times $h(x)$ at infinity wraps around the group.

Considering all gauge configurations of a given topology, one can
search for one that minimizes the action.  Being a local minimum, this
is also automatically a solution to the Yang-Mills equations of
motion.  For a single wrapping, this is the famous ``instanton''
solution.  Locating the action peak at the origin, this takes the form
\begin{equation}
A_\mu={-ix^2\over g(x^2+\rho^2)} h^\dagger \partial_\mu h
\end{equation}
where we take 
\begin{equation}
h(x_\mu)={t+i\vec\tau\cdot\vec x \over \sqrt{\vec x^2+t^2}}\in SU(2).
\end{equation}
Note how the factor $x^2\over x^2+\rho^2$ mollifies the singularity in
$h^\dagger \partial_\mu h$ at the origin.

This solution depends on an arbitrary parameter $\rho$ which
characterizes the instanton size. Indeed, there must be such a scale
parameter since the pure classical gauge theory is scale invariant.
Integrating over space gives the classical instanton action 
\begin{equation}
\label{classical}
S_I={1\over 2}\int d^4x\ {\rm Tr}\ F_{\mu\nu}F_{\mu\nu}=
{8\pi^2\over g^2}.
\end{equation}
Because of the inverse coupling dependence of this action, any physics
associated with non-trivial topology is necessarily non-perturbative.

\section{Quantum issues}
\label{quantum}

Of course for particle physics one is interested in the quantum
theory.  The usual procedure for this is to introduce a path
integral involving a sum over all possible configurations of the gauge
fields.  There are several subtle and non-intuitive issues that arise
upon quantization.  In general the winding number
\begin{equation}
\nu={g^2\over 16\pi^2}\int d^4x\ {\rm Tr}F\tilde F
\end{equation} 
is robust under smooth field deformations.  However in path integrals
it is well known that typical paths are non-differentiable. On such it
is not clear if the concept of topology is well defined.  In the next
subsections I discuss some of the subtleties that arise.

\subsection{Positivity and the topological susceptibility}

A natural quantity to study in the quantum theory is
the topological susceptibility, defined as 
\begin{equation}
\label{suscept}
\xi= \langle {\nu^2\over V} \rangle = \left( {g^2\over 16\pi^2}\right
)^2
\int d^4x \langle F\tilde F(x)\  F\tilde F(0)\rangle
\end{equation}
This directly measures the typical quantum fluctuations in the winding
number, and since it is the expectation of a square, it must be
positive.\footnote{This is in the absence of fermions.  With such
present, a negative fermion determinant can give rise to a
susceptibility that is formally negative \cite{Creutz:2013xfa}.} But
the combination $F\tilde F$ as an operator is odd under time reversal.
By reflection positivity \cite{Seiler:2001je}, this means that for any
non-zero separation $x$ one must have
\begin{equation}
\langle F\tilde F(x)\   F\tilde F(0)\rangle <0
\end{equation}
Thus the integral in Eq.~(\ref{suscept}) receives a negative
contribution from all $x\ne 0$.  In order to obtain a positive
susceptibility, there must be a positive ``contact'' term from a
singularity at $x=0$.  This observation emphasizes that long and short
distance phenomena are intimately entwined when non-perturbative effects from
topology are important.  It is inherently dangerous to try to separate
out perturbative effects by discussing only short distances.  This
point will arise again when I discuss the theory with quarks
present. 

\subsection{The interplay of topology and asymptotic freedom}

The well known phenomenon of asymptotic freedom
\cite{Politzer:1973fx,Gross:1973id,Gross:1973ju,
Caswell:1974gg,Jones:1974mm} implies that the effective gauge coupling
$g$ depends on scale and decreases at short distances.  On
quantization the conformal invariance of the classical theory is lost
and a scale enters the problem through what is called ``dimensional
transmutation'' \cite{Coleman:1973jx}. For the lattice theory this
means that the bare coupling at the lattice scale, $g(a)$, must be
taken to zero for the continuum limit.  In effect, the lattice spacing
$a$ represents an ultraviolet cutoff on the theory.  This decrease in
the coupling is logarithmic
\begin{equation}
g^2(a)\sim {1\over 2\beta_0 \log(1/\Lambda a)}
\end{equation}
where $\Lambda=\Lambda_{qcd}$ is an integration constant from the
renormalization group equation.  This is a dimensional parameter that
sets the scale for particle masses in the quantum theory.  More
precisely, $\Lambda$ satisfies
\begin{equation}
\Lambda_{qcd}= {1\over a}e^{1/2\beta_0g^2}
g^{-\beta_1/\beta_0^2}
(1+O(g^2))\qquad {\longrightarrow_{a\rightarrow 0}}\qquad {\rm constant}
\end{equation}
where
\begin{equation}
\beta_0={1\over 16\pi^2}(11-2N_f/3),\qquad
\beta_1=\left({1\over 16\pi^2}\right)^2(102-22N_f/3).
\end{equation}
Here $N_f$ denotes the number of quark species.

Because the coupling goes to zero, small classical instantons are
suppressed in the continuum limit.  Naively combining the classical
action in Eq.~(\ref{classical}) with the asymptotic freedom result
gives
\begin{equation}
\label{semiclassical}
e^{-S_I}\sim e^{-8\pi^2/g(a)^2} \sim e^{-16\pi^2 \beta_0\log(a\Lambda)}
\sim a^{11-2N_f/3}.
\end{equation}
This represents a strong suppression of instantons by a power law in
the lattice spacing.  For 3 flavors, this is by $a^{9}$ per unit
lattice volume, or $a^{5}$ per unit physical volume.  Because of this
it is sometimes argued that one can ignore instantons at small lattice
spacing.

This is a wrong conclusion. As will be discussed later, topological
excitations directly affect the hadron spectrum in the continuum
limit.  In other words, they contribute to physics at order $a^0$.
This difference arises because the above semi-classical argument
ignores quantum fluctuations.  Typical quantum paths are far from
smooth, indeed, they are non-differentiable.  The most direct way to
see how topology modifies the hadron spectrum is through their
influence on the quark fields, to which I now turn.

\section{Fermions and the index theorem}
\label{index}

The interactions of the quarks with the gluon fields enter through the
Dirac contribution to the action $S_f=\overline\psi (D+m)\psi$.  In
practice the kinetic part of the Dirac operator $D$ satisfies
gamma-five hermiticity,
\begin{equation}
D^\dagger=-D=\gamma_5 D \gamma_5.
\end{equation}
In continuum discussions $D$ is also anti-Hermitian and anti-commutes
with $\gamma_5$.  This latter condition is usually modified on the
lattice, but that won't concern us for the moment.

A particularly important consequence of topology is the index
theorem.  This says that when the gauge field has non-trivial winding,
the Dirac operator $D$ will have exact zero modes, i.e. there exist
functions that satisfy
\begin{equation}
D |\psi\rangle=0.
\end{equation}
Furthermore, on the space of zero modes $\gamma_5$ can be
diagonalized.  Thus these modes can be considered to be chiral
\begin{equation}
\gamma_5|\psi\rangle=\pm |\psi\rangle.
\end{equation}
The theorem states that the winding number equals the difference
between the number of right and left handed modes,
\begin{equation}
\nu=n_+-n_-.
\end{equation}

The index theorem implies that on a fixed gauge field configuration
${\rm Tr} \gamma_5=\nu$.  At first sight this seems to be a strange
result since, when thought of as a four by four matrix, ${\rm
Tr}\ \gamma_5=0$.  But, as argued by Fujikawa,
\cite{Fujikawa:1979ay} this naive conclusion must be modified in a
regulated theory.  To see this, it is natural to use the eigenstates
of $D$ to define the trace.  Considering a complete set of eigenstates
\begin{equation}
D|\psi_i\rangle=\lambda_i|\psi_i\rangle,
\end{equation}
a natural definition for the trace is
\begin{equation}
{\rm Tr}\gamma_5=\sum_i  \langle\psi_i|\gamma_5|\psi_i\rangle
\end{equation}
All non-zero eigenstates occur in chiral pairs; for an eigenstate
$|\psi_i\rangle$, then
\begin{equation}
D\gamma_5 |\psi\rangle=-\lambda\gamma_5 |\psi\rangle
=\lambda^*\gamma_5|\psi\rangle 
\end{equation}
From this $|\psi\rangle$ and $|\gamma_5\psi\rangle$ are orthogonal
when $\lambda\ne 0$, and in turn the space spanned by $|\psi\rangle$
and $|\gamma_5 \psi\rangle$ gives no contribution to ${\rm
Tr}\gamma_5$.  Only the zero modes count towards the trace and one has
the basic result
\begin{equation}
{\rm Tr}\gamma_5=\sum_i \langle \psi_i |\gamma_5|\psi_i\rangle=\nu.
\end{equation}

Since $\gamma_5$ arises from a traceless four by four matrix, it is
natural to ask what happened to the opposite chirality states? In a
continuum discussion it is easiest to think of them as being lost at
``infinity,'' i.e. they are at an energy beyond the cutoff.  On the
lattice there is no real infinity, and the answer depends on the
details of the fermionic action.  With Wilson
fermions \cite{Wilson:1975id}, the would-be zero modes can acquire a
small real part and compensating real eigenvalues appear in the
doubler region.  Including all states, $\gamma_5$ remains traceless.
With overlap fermions \cite{Narayanan:1993sk,Neuberger:1997fp}, all
eigenvalues of the Dirac operator lie on a circle; for every zero mode
there is a compensating real eigenvalue with opposite chirality on the
far side of the circle.

This brings us back to the earlier conclusion that this phenomena
responsible for the anomaly involves both long and short distances.
Even for large instantons, there is a compensating mode which is at or
beyond the cutoff region.  And for small instantons, they can combine
with these other states and ``fall through the lattice.''  The details
depend on the specific cutoff in place; indeed, these can be scheme
and scale dependent.

The way the zero modes from topology bring about the anomaly can be
nicely understood in terms of the fermionic measure
\cite{Fujikawa:1979ay}.  In particular, the fact that $\gamma_5$ is
formally not traceless means that the change of variables
\begin{equation}
\psi \rightarrow e^{i\gamma_5\theta}\psi
\end{equation} 
changes the fermion measure in the path integral
\begin{equation}
(d\psi d\overline\psi)\rightarrow e^{i\theta {\rm Tr}\gamma_5}
(d\psi d\overline\psi)=e^{i\nu\theta}(d\psi d\overline\psi).
\end{equation}
Thus such a change in variables inserts a factor of $e^{i\nu\theta}$
into the path integral weight.  This gives rise to an inequivalent
theory.  Starting with the naive path integral, this rotation gives
what is often called the ``Theta vacuum,'' the ground state of an
independent and physically distinct theory in which CP symmetry is
explicitly broken.

\subsection{Fixed topology}

Because the space of smooth fields breaks up into distinct sectors,
it turns out that in a simulation tunneling between these sectors is
difficult.  This gives rise to long correlation times.  This issue
becomes even more severe when the quarks become light.  In this case
the the near zero modes suppress configurations of non-trivial
topology.

This raises the question of what would happen if one were to ignore
this tunneling and work in a sector of fixed total topology.  This can
be implemented formally by integrating over $\Theta$ to select out
the sector of interest.  Thus consider the fixed topology path
integral
\begin{equation}
\label{znu}
Z_{\nu}(m_q)=\int {d\Theta\over 2\pi} \ e^{i\nu\Theta}\ Z(m_q,\Theta)
\end{equation} 
In some ways this seems like a rather perverse thing to do since each
Theta vacuum represents a physically different theory.  Nevertheless
it has been argued \cite{Aoki:2007ka} that as the system volume becomes
large one can still obtain valid physics.
As the volume increases the path integral grows/decreases
exponentially in the free energy of the four dimensional statistical
system being simulated
\begin{equation}
Z(m_q,\Theta)
=\int (dA)(d\overline\psi)(d\psi)\ e^{-S(m_q,\Theta)}\  =\  e^{-VF(m_q,\Theta)}
\end{equation}
Although each value of $(m_q, \Theta)$ represents an physically
different field theory, as $V\rightarrow \infty$ with fixed $\nu$ the
integral over $\Theta$ in Eq.~(\ref{znu}) will be dominated by the
saddle point at $\Theta=0$.  In physical terms, at large enough volume
a few instantons can ``hide behind the moon.''  The conclusion is that
at large volume, despite the long correlation time in the total
winding number, local observables are presumably much better behaved.

\section{The pseudo-scalar spectrum}
\label{pseudoscalars}

Possibly the most direct physical consequences of topology appear in
the spectrum of the pseudo-scalar mesons.  Consider two flavor QCD
with light but non-degenerate quark masses.  As usual, label the quark
fields as $u$ and $d$.  There are four natural
pseudo-scalar bilinears in the quark fields
\begin{eqnarray}
&i\overline u\gamma_5 u\cr
&i\overline d\gamma_5 d\cr
&i\overline u\gamma_5 d\sim\pi_+\cr
&i\overline d\gamma_5 u\sim\pi_-\cr
% &F_{\mu\nu}\tilde F_{\mu\nu}\cr
\end{eqnarray}
All of these combinations involve a helicity flip, for example
$
i\overline u\gamma_5 u
=i\overline u_L\gamma_5 u_R+i\overline u_R\gamma_5 u_L.
$
Now a well known property of gauge theories is the suppression of
helicity flip processes in the chiral limit.  This naively suggests
that the mixing of  $i\overline u\gamma_5 u$ with $i\overline
d\gamma_5 d$ should be suppressed by a factor of $m_um_d$.  Without
such mixing there should be two light neutral pseudo-scalar pions, one
primarily made of up quarks and the second from down quarks.

This is of course wrong.  There is only one neutral pion, not two.  It
is the anomaly that strongly couples the up and down combinations
through what is commonly called the effective ``t'Hooft
vertex'' \cite{'tHooft:1976fv}. In this way the symmetric combination
\begin{equation}
\eta^\prime\sim\overline u\gamma_5 u + \overline d\gamma_5 d
\end{equation}
is not a pseudo-Goldstone boson and acquires a mass of order the QCD
scale
\begin{equation}
M_{\eta^\prime}\propto \Lambda_{qcd}+O(m_u,m_d).
\end{equation}

The mixing responsible for the eta prime mass leaves behind the orthogonal
combination
\begin{equation}
\pi_0 \sim i\overline u\gamma_5 u - i\overline d\gamma_5 d,
\end{equation}
which is, of course, the neutral pion.  In the process, isospin
breaking is suppressed to a higher order in the chiral expansion
\begin{equation}
M_{\pi_0}^2=M_{\pi_\pm}^2-O((m_u-m_d)^2).
\end{equation}

The basic result is that the $\eta^\prime$ meson is not a Goldstone
boson because 
\begin{equation}
\psi \rightarrow e^{i\gamma_5\theta}\psi
\end{equation}
is not a symmetry of the quantum theory.  In the absence of such a
symmetry, the mass of the $\eta^\prime$ is proportional to
$\Lambda_{qcd}$, the scale of the strong interactions, and does not
vanish as the quark masses go to zero
\begin{equation}
M_{\eta^\prime}^2 
\propto {1\over a^2} e^{1/\beta_0g^2} g^{-\beta_1/\beta_0^2}(1+O(g^2)\sim
a^0.
\end{equation}
This behavior was alluded to earlier and is a direct indication of
how the semi-classical estimate from Eq.~(\ref{semiclassical})
substantially understates the importance of
topology in the quantum theory.

\section{Isospin breaking and quark masses}
\label{isosec}

\begin{figure}
  \includegraphics[height=.5\textheight]{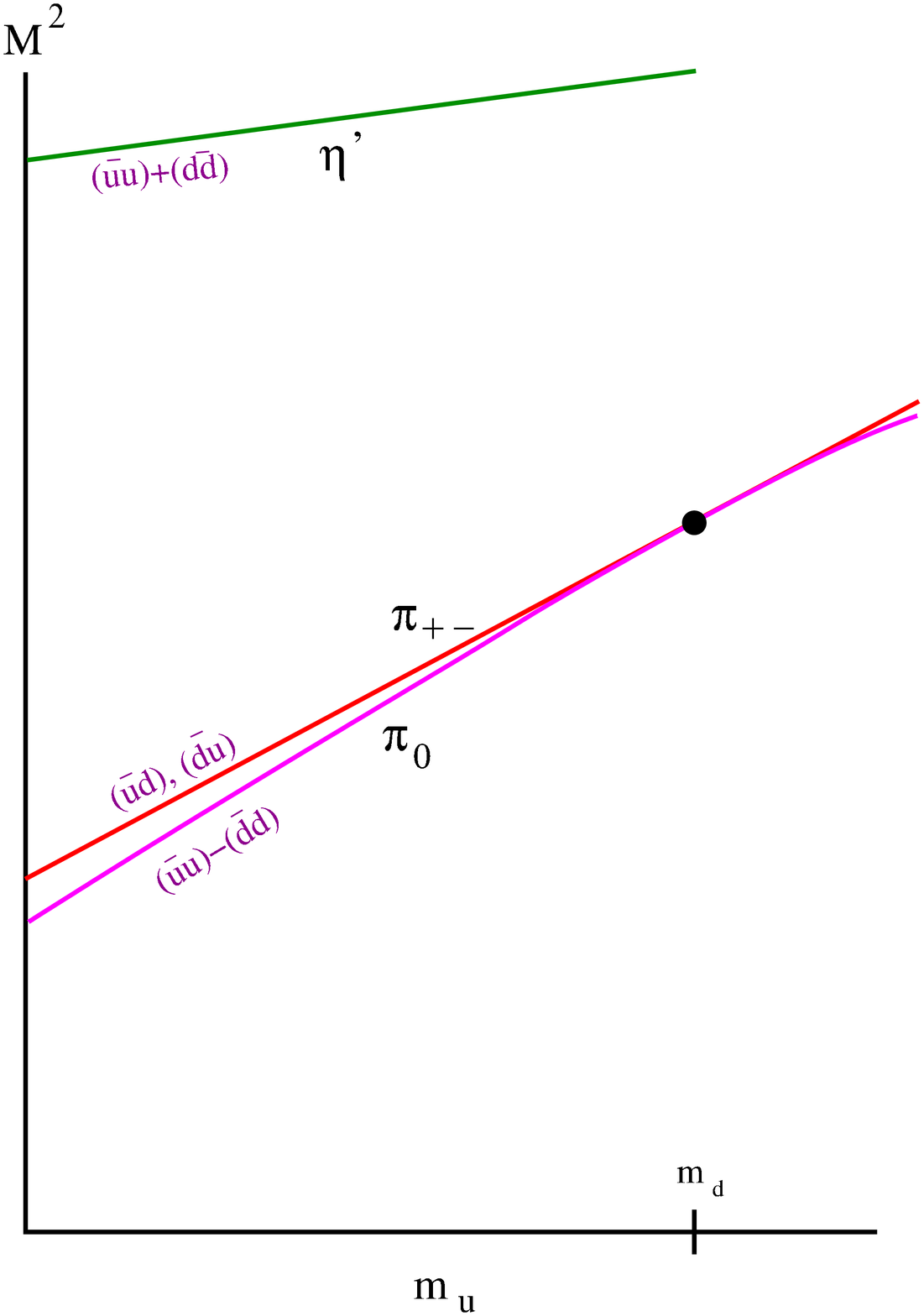}
  \caption{A mass gap will persist when the up quark mass vanishes but
    the down quark remains massive.  Higher order effects split the
  charged and neutral pions.}
\label{iso1}
\end{figure}

I now turn to some interesting properties of the theory as a function
of the up quark mass $m_u$ when the down quark mass $m_d$ is fixed at
a non-zero value.  Consider the situation where both are light
compared to the strong scale, Then chiral symmetry predicts
\begin{equation}
M_\pi^2 \propto {m_u+m_d\over 2}+O(m_q^2).
\end{equation}
As discussed above, the eta prime remains massive with $M_{\eta'}
\sim \Lambda_{qcd}$.  But an important observation is that a finite mass gap
remains if only the up quark is massless.  This is sketched in
Fig. \ref{iso1}.

The effect of isospin breaking on the pion masses is of higher order
in the quark masses\footnote{I ignore the splitting coming from
electromagnetism, which dominates the physical mass difference.  In
the following I explore the hypothetical situation where the mass
splitting is large and dominates the electromagnetic part.}
\begin{equation}
M_{\pi_\pm}^2-M_{\pi_0}^2 \propto (m_d-m_u)^2,
\end{equation}
At this quadratic order in the quark mass difference, isospin breaking
is expected to induce some mixing between the neutral pion, the eta
prime, and pseudo-scalar glueballs.  This mixing is expected to make
the neutral pion become the lightest of the three pions.

\subsection{The Dashen phase}

\begin{figure}
  \includegraphics[height=.4\textheight]{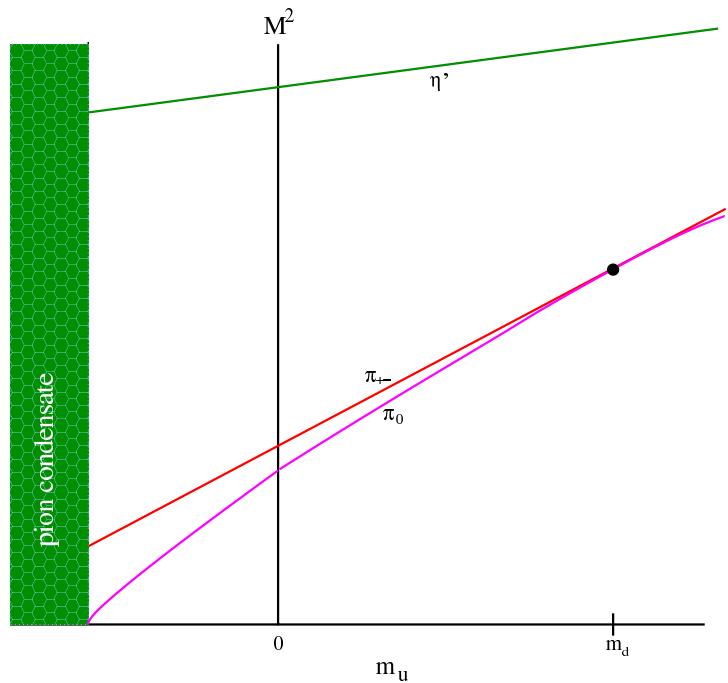} \caption{No
  singularity is expected as the up quark mass passes through zero at
  fixed down quark mass.  On the other hand, for sufficiently negative
  up quark mass the neutral pion can condense into a CP violating
  phase.}
\label{iso2}
\end{figure}

Since there is a mass gap at vanishing up quark mass, it is natural to
ask what would happen if the up quark mass were to become negative.
In effective chiral Lagrangian models, both linear and nonlinear
\cite{Creutz:2011hy,Aoki:2014moa,Horkel:2014nha,Horkel:2014lta}, 
the spectrum remains well behaved, without any singularity on passing
through the zero mass point.  There is, however, a natural limit to
the size of the mass splitting when the mixing with the eta prime is
large enough to make the neutral pion massless.  Beyond that point the
neutral pion field acquires an expectation value and one finds a new
phase in which CP is spontaneously broken.  Because the product of the
quark masses is negative, this occurs in a region where the parameter
$\Theta$ formally takes the value $\pi$.  The possibility of such a
behavior was conjectured some time ago by Dashen \cite{Dashen:1970et}.
This behavior is qualitatively sketched in Fig. \ref{iso2}.

\begin{figure}
  \includegraphics[height=.32\textheight]{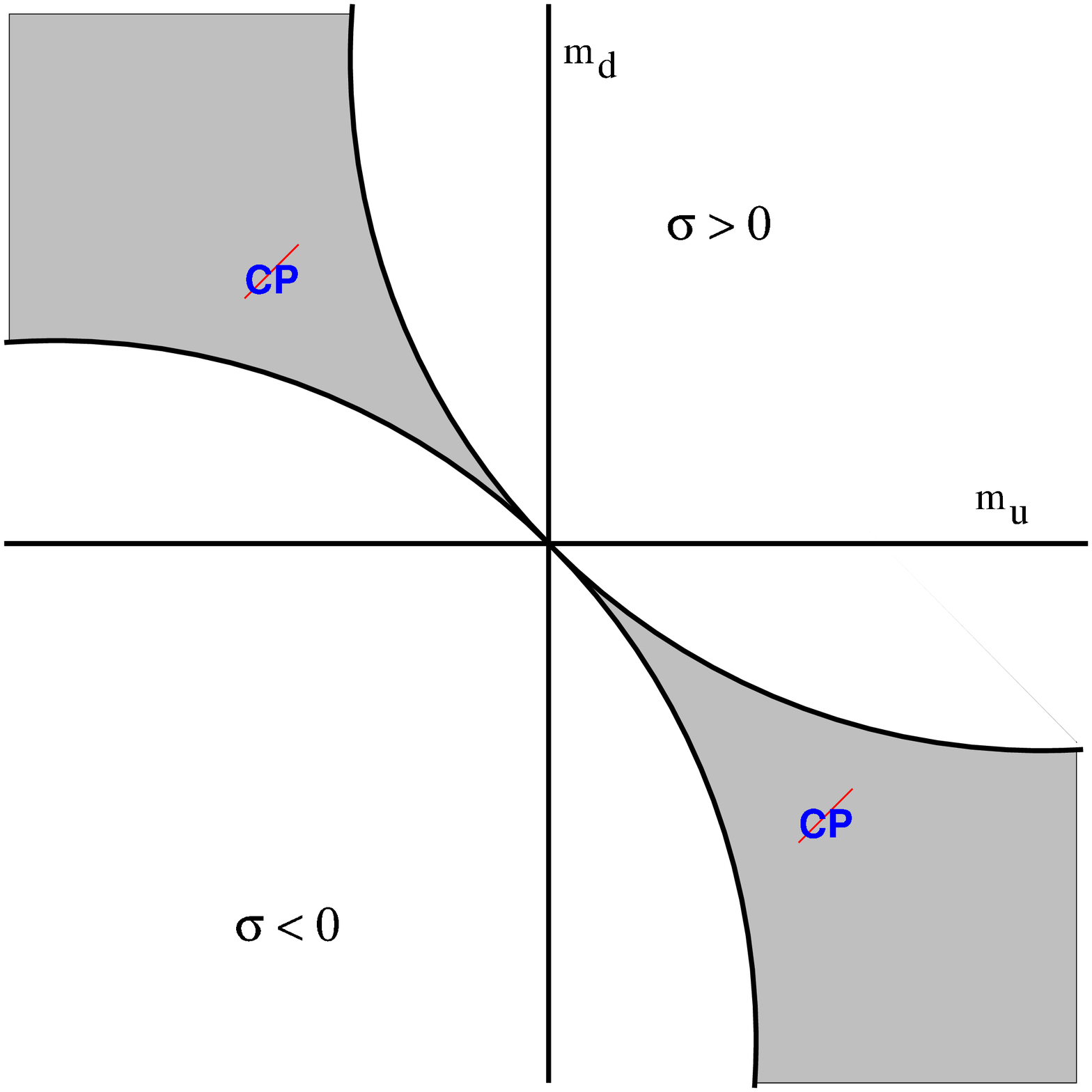}
  \caption{The qualitative phase diagram for two flavor QCD as a
    function of the up and down quark masses.  The CP violating Dashen
    phase occurs in the region where the two masses have opposite
    sign.}
\label{iso4}
\end{figure}

In the effective Lagrangian models this is an Ising like transition
occurring at a negative up-quark mass.  The order parameter is the
expectation value of the neutral pion, $\langle \pi_0 \rangle \ne 0$.
Because the pion is CP odd, this symmetry is spontaneously broken in
the Dashen phase.  The qualitative phase diagram as a function of the
two quark masses is sketched in Fig. \ref{iso4}.  This structure gives
rise to several remarkable conclusions.  First, a mass gap persists at
{$m_u=0$} when {$m_d\ne 0$}.  This means that despite the fact that
the Dirac operator can have small eigenvalues, all long distance
physics is exponentially suppressed by this mass gap.  Second, the
presence of a second order transition when neither $m_u$ nor $m_d$
vanishes shows that it is possible to have a divergent correlation
length and long distance physics in a regime where the Dirac operator
has no small eigenvalues.

Fig.~\ref{iso4} also illustrates important symmetries.  Note the
reflection symmetry about the 45 degree $m_u+m_d$ direction, and
another about the $m_u-m_d$ direction.  These symmetries correspond to
an isoscalar mass term $m_u+m_d$ and an isovector mass $m_u-m_d$.
Each of these are separately protected from any additive
renormalization.  However there is no symmetry between these possible
mass terms.  Because of this, the up and down quark masses are not
individually protected.  In particular non-perturbative mass
renormalization does not maintain the perturbative property of being
``flavor blind.''

\begin{figure}
  \includegraphics[height=.175\textheight]{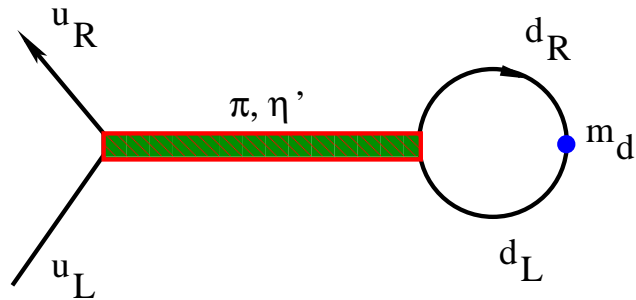}
  \caption{A small down quark mass can induce an effective up quark
    mass through intermediate meson exchange, even if the up quark is
    perturbatively massless.  
}
\label{induced2}
\end{figure}

This lack of symmetry in the individual quark masses arises because
non-perturbative contributions can mix quark masses.
Fig.~\ref{induced2} shows one way to think of this mixing.  The eta
prime and neutral pion mesons are both non-trivial mixtures of
$\overline u u$ and $\overline d d$ quark-antiquark pairs.  Because
these mesons are non-degenerate, their contribution to the mixing of
$\overline u u$ and $\overline d d$ cannot cancel.  Thus a small down
quark mass will induce an effective up quark mass, even if the up
quark is massless in perturbation theory.  This induced up quark mass
is of form
\begin{equation}
\delta m_u \propto {(M_{\eta^\prime}-M_\pi)\over\Lambda_{qcd}}m_d.
\end{equation}
For three flavors there is also a contribution from the strange quark
giving $\delta m_u\propto {m_dm_s \over m_d+m_s}$.  Note that this is the
same form as the Kaplan and Manohar ambiguity in chiral
perturbation theory \cite{Kaplan:1986ru}.

This discussion shows that the concept of a single massless quark is
not renormalization group invariant.  Indeed, is
there any experimental way to tell if $m_u=0$?  It is often stated
that the strong CP problem would be solved  if the up quark mass
vanishes.  But can this make any sense if such a concept is
ill-defined? This cannot be answered in the context of the
$\overline{MS}$ scheme since that is perturbative and these effects
are purely non-perturbative.

\section{The lattice}
\label{latticesec}

So, a non-perturbative approach is necessary to understand the quark
masses.  This leads us directly to the lattice.  Naively all one needs
to do is adjust the lattice parameters to get the hadron spectrum, and
then read off the quark masses to answer such questions as whether
$m_u=0$.

But this leaves many open questions.  There are many different lattice
formulations.  Are the quark masses unique between them?  What defines
the quark mass anyway?  One could look for poles in the quark
propagator, but the propagator is gauge dependent and the result might
depend on the gauge chosen.  These are all non-trivial questions that
have not been fully answered.

\begin{figure}
  \includegraphics[height=.32\textheight]{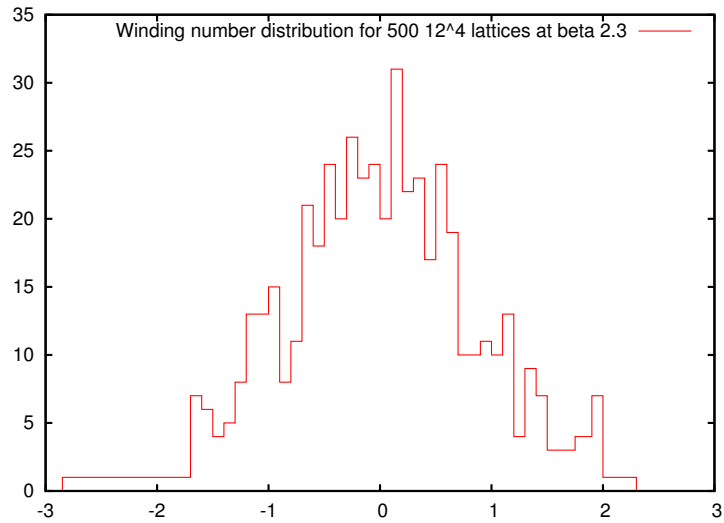}
  \caption{The distribution of topological charge on an ensemble of
    SU(2) gauge theory lattices as constructed from a set of loops
    around the lattice hyper-cubes \cite{Creutz:2010ec}.  Without some sort
    of smoothing the distribution does not show peaks at integer values.
}
\label{fig1}
\end{figure}

The quark mass is closely connected with topology; if $m_u=0$ the
topological susceptibility should vanish.  But how does one define
topology on a discrete lattice?  Small instantons can ``fall through
the lattice'' and the concept of topology is lost at the outset.  One
can construct a combination of loops around a hyper-cube that reduces
to $F\tilde F$ in the naive continuum limit.  There are a variety of
ways to do this, but the resulting topological charge is not generally
an integer.  One example is shown in Fig.~\ref{fig1}, taken
from \cite{Creutz:2010ec}.

\begin{figure}
  \includegraphics[height=.45\textheight,angle=-90]{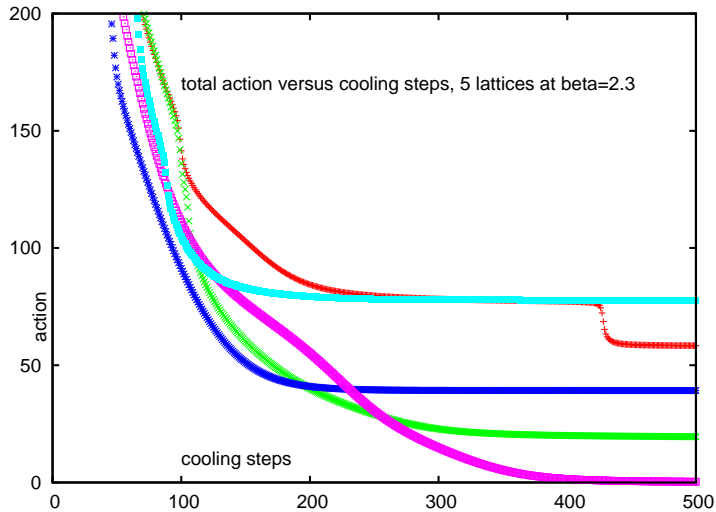}
  \caption{
  The lattice action for five SU(2) lattices as  cooling removes
  dislocations.  The action settles to multiples of the classical
  instanton value.
}
\label{actioncooling}
\end{figure}

The topological charge of individual configurations can be driven to
integers by various cooling algorithms.  These remove rough
configurations and the action settles into multiples of the classical
instanton result.  The result of such a cooling process is shown in
Fig.~\ref{actioncooling}, taken from Ref.~\cite{Creutz:2010ec}.  Over
the years there have been many such studies with a variety of
methods \cite{Teper:1985rb, deForcrand:1997ut, DelDebbio:2004ns,
Bruckmann:2006wf,Cichy:2014qta}.

\begin{figure}
  \includegraphics[height=.45\textheight,angle=-90]{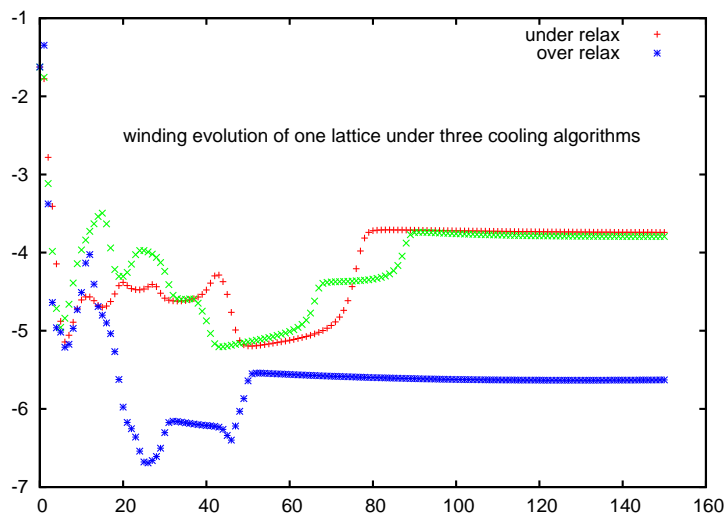}
  \caption{
The evolution of the action for a single lattice under different
cooling algorithms.  The result can depend on the details of the
algorithm as well as the time of cooling.
}
\label{fig12}
\end{figure}

Cooling often gives a stable result, but ambiguous cases do appear.
Indeed, the final winding can depend on details of the cooling
algorithm.  For example, Fig.~\ref{fig12}, also taken from
Ref.~\cite{Creutz:2010ec}, shows the evolution of a single
configuration under several rather different cooling algorithms.  This
leaves a variety of open questions.  With which gauge action should
one cool, particularly when dynamical fermions are present?  How long
should one cool?  With too much cooling, will small ``instantons''
eventually collapse?

At this point I remark on a rather technical point.  If one puts a
bound on how far any individual plaquette can be from the identity,
then it is possible to implement a unique interpolation of the gauge
fields through the lattice hyper-cubes.  If this ``admissibility
condition'' is imposed, the winding number becomes well defined.
Specifically, Ref.~\cite{Luscher:1981zq} has shown that if the trace
if each plaquette in an SU(3) gauge theory takes a value $P<\sim .03$,
then instantons can no longer collapse and the configuration has a
unique winding number.

The problem with the admissibility constraint is that it requires a
non-Hermitian Hamiltonian.  The transfer matrix relates the path
integral to the Hamiltonian through the transfer matrix 
\begin{equation}
Z={\rm Tr} e^{-\beta H}={\rm Tr} (e^{-aH})^{N_t}.
\end{equation}
A Hermitian $H$ requires $\langle \psi|e^{-aH}|\psi\rangle >0$ for
every state $\psi$.  It can be shown \cite{Creutz:2004ir} that this
requires the plaquette weight to be analytic over the gauge group.
That in itself is inconsistent with the admissibility constraint.

Can the index theorem provide another approach to defining the
topological charge?  For example, count the small real eigenvalues of
the Wilson operator and use the result as a definition of the charge?
This becomes tricky since at finite cutoff these are not exact zeros
but are spread over a region of the real axis.  Thus the word
``small'' is a bit arbitrary.  The unresolved question is whether the
eigenvalue distribution in the first Wilson ``circle'' goes to zero
fast enough to remove this ambiguity.

One might consider the zero modes of the overlap operator, which are
indeed exact zero modes.  The problem is that the overlap operator is
not unique, and depends on the chosen ``domain wall height.''  The
uniqueness of this again relies on the density of eigenvalues in the
first Wilson ``circle'' going to zero sufficiently rapidly.

\subsection{Should we care?}

Should we care if there is a small ambiguity in defining topology?
After all, this is not something directly measured in laboratory
experiments.  It is perhaps better to concentrated on something like
the mass of the eta prime, which is clearly physical. Of course there
is the Witten-Veneziano formula \cite{Witten:1979vv,Veneziano:1979ec}
relating this to topology, but that is valid only in the large $N_c$
limit while $N_c=3$ for physics.

As discussed above, topology is closely related to the issue of
whether $m_u=0$ or not.  Fig.~\ref{iso4} shows that there is no
symmetry around the $m_d$ axis.  Is there perhaps some Ward identity
that fixes the location of this line?  Any such relation involves
anomalous currents and thus must bring in the topological
susceptibility.  Thus any ambiguities in defining a vanishing quark
mass or the topological susceptibility are directly coupled.  And one
should remember from \cite{Kaplan:1986ru} that these ideas are already
ambiguous at the level of chiral perturbation theory.

\section{Summary}
\label{summary}

I hope I have convinced you that the role of topology in gauge
theories is a rich and fascinating topic.  There are important
consequences for understanding the light hadron spectrum, and thus the
topic is highly relevant to this meeting.  It is also important to
remember that conventional perturbation theory misses many of these
issues, for example the mass mixing effects between species.

\blockcomment
Several of these ideas have proven to be controversial.  The issue that
$m_u=0$ may be scheme dependent is not widely accepted, nor is the
ambiguity in topological susceptibility.  A directly related
controversy is over the failure of staggered fermions to reproduce
properly the anomaly after the rooting trick is
applied \cite{Creutz:2007yg,Creutz:2007rk}.
\endcomment

I have also discussed the fascinating phase structure that can appear
with negative mass quarks.  This includes a phase with pion
condensation and spontaneous CP violation.  While this is not directly
relevant for the usual strong interactions, it might be interesting to
consider such a mechanism to introduce CP violation in models of
unification through a new strong dynamics.

\begin{theacknowledgments}
I'm grateful for an Alexander von Humboldt Research Award which
    partially supported my attendance at this meeting.
\end{theacknowledgments}

\end{document}